\documentclass[useAMS,usenatbib]{mn2e}
\usepackage{natbib}
\usepackage{graphicx}
\usepackage{txfonts}

\begin{document}
\title[Spiral arms of the Milky Way]{Offset between stellar spiral arms and
  gas arms of the Milky Way}

\author[L. G. Hou and J. L. Han]
{L.~G. Hou\thanks{E-mail:
    lghou@nao.cas.cn} and J. L. Han
\\ National Astronomical
  Observatories, Chinese Academy of Sciences, 20A DaTun Road, ChaoYang
  District, Beijing 100012, PR China }

\date{Accepted 2015 ... Received 2015 ...}

\pagerange{\pageref{firstpage}--\pageref{lastpage}} \pubyear{2015}

\maketitle

\label{firstpage}

\begin{abstract}
Spiral arms shown by different components may not be spatially
coincident, which can constrain formation mechanisms of spiral
structure in a galaxy. We reassess the spiral arm tangency directions
in the Milky Way through identifying the bump features in the
longitude plots of survey data for infrared stars, radio recombination
lines (RRLs), star formation sites, CO, high density regions in
clouds, and HI. The bump peaks are taken as indications for arm
tangencies, which are close to the real density peaks near the spiral
arm tangency point but often have $\sim$ 1$^\circ$ offset to the
interior of spiral arms. The arm tangencies identified from the
longitudes plots for RRLs, HII regions, methanol masers, CO, high
density gas regions, and HI gas appear nearly the same Galactic
longitude, and therefore there is no obvious offset for spiral arms
traced by different gas components. However, we find obvious
displacements of 1.3$^\circ-$ 5.8$^\circ$ between gaseous bump peaks
from the directions of the maximum density of old stars near the
tangencies of the Scutum-Centaurus Arm, the northern part of the Near
3 kpc Arm, and maybe also the Sagittarius Arm. The offsets between the
density peaks of gas and old stars for spiral arms are comparable with
the arm widths, which is consistent with expectations for
  quasi-stationary density wave in our Galaxy.
\end{abstract}


%

\section{Introduction}
\label{sect:intro}

The formation and evolution of spiral arms in spiral galaxies is a
fundamental problem in astronomy for more than 90 years
\citep[e.g.,][]{hub26,lin27}. Observationally, spiral galaxies can be
classified into, for example by \citet{elm90}, flocculent (e.g.,
NGC\,4414), multi-armed (e.g., M\,101), and grand design galaxies
(e.g., M\,51). The mechanisms for spiral formation \citep[see][for a
  review]{db14} could be (1) quasi-stationary density wave theory
\citep[e.g.,][]{ls64,ls66}; (2) localized instabilities,
perturbations, or noise induced kinematic spirals
\citep[e.g.,][]{sc84}; (3) dynamically tidal interactions
\citep[e.g.,][]{tt72}. It is difficult to identify which mechanism
dominates spiral structure in a specific galaxy
\citep[e.g.,][]{bt08}. Careful comparisons of spiral features between
observations and theories are necessary to distinguish these
mechanisms \citep[][]{frb+11}.

\begin{figure}
\centering\includegraphics[width=0.45\textwidth]{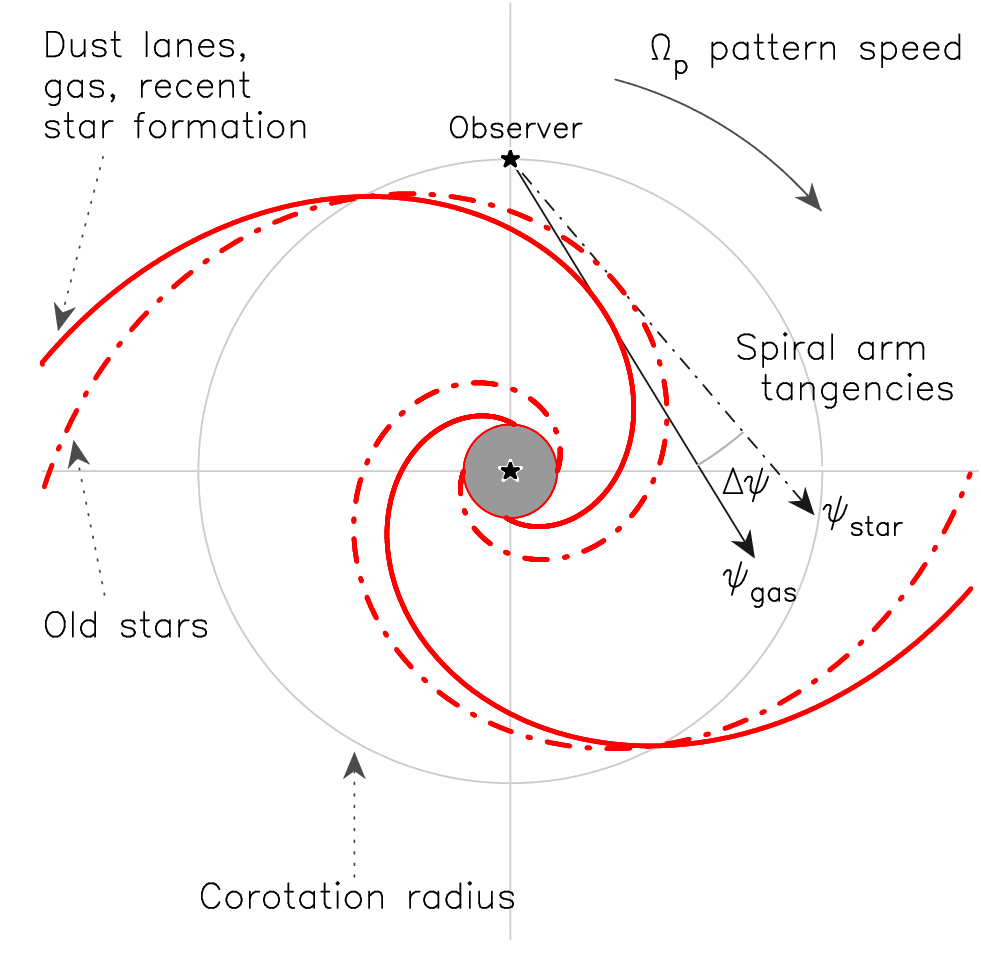}\\
\caption{Schematic of the relative position between gas and old star
  components in spiral arms according to a quasi-stationary spiral
  structure in a face-on view [see also Fig. 1 of \citet{mgb09},
    Fig. 2 of \citet{frb+11}, and Fig. 7 of \citet{rob69}]. Spiral arm
  tangencies for gas ($\psi_{gas}$) and stars ($\psi_{star}$) are
  shown by the density peaks near the tangency points of ideal spiral
  arms. Thus, the difference $\Delta \psi=|\psi_{star} - \psi_{gas}|$
  can be used to indicates the possible spatial offset of different
  components in spiral arms.}
\label{show}
\end{figure}

The quasi-stationary density wave theory \citep{rob69} predicts a
spatial offset of different arm components, e.g., old stars, star
forming regions, molecular gas and atomic gas, due to possible delay
in star formation from gas gathering (see Fig~.\ref{show}). The other
mechanisms do not predict the offset between the stellar spiral arms
and gaseous arms \citep[see, e.g.,][]{db14,bme15}. Face-on spiral
galaxies (e.g., M\,51, M\,81) are the ideal objects to test these
mechanisms. Many observations have been made
\citep[e.g.,][]{mvb72,rot75,trw+08,eks+09,frb+11,mp14}, but different
conclusions were made by different authors even from the high quality
data of HI, CO, 24~$\mu$m infrared and UV images of the same sample of
galaxies, mainly because of large uncertainties in measuring the
spiral arm properties \citep[][]{db14}.

For the Milky Way Galaxy, the possible spatial offset of different
spiral arm components is very difficult to measure and far from clear
at present, mainly because the spiral pattern and positions of arms
\citep[e.g.][]{val08,fc10,hh14,pdap14,car15} have not been well
determined. However, the directions of spiral arm tangencies in our
Galaxy can be estimated \citep[e.g.,][]{ben09}, which can be used to
check the possible offset between different components of spiral arms
(e.g., see Fig.~\ref{show}).

Previously spiral arm tangencies have been identified from the
Galactic plane surveys of HI \citep[e.g.,][]{bs70,wea70}, $^{12}$CO
\citep[e.g.,][]{sss85,ssr85,dect86,gcb+87,bca+88,bact89,bcmn00,bml00,dt11},
$^{13}$CO \citep[e.g.,][]{sl06}, radio continuum emission at 408 MHz
\citep[e.g.,][]{bkb85} and at 86 MHz \citep[e.g.,][]{mhs58}, HII
regions \citep[e.g.,][]{loc79,loc89,dwbw80,hh14}, near infrared
emission \citep[e.g.,][]{hmm+81}, far infrared dust emission
\citep[e.g.,][]{bdt90,dri00}, far infrared cooling lines
\citep{swh10}, 870 $\mu$m continuum \citep[][]{btl+12}, 6.7-GHz
methanol masers \citep[e.g.,][]{cfg+11,gcm+11,gcf+12}, and old stars
\citep{cbm+09} for the Scutum Arm, the Sagittarius Arm, the Carina
Arm, the Centaurus Arm and the Norma Arm \citep[e.g., see Table~1
  of][and {Vall{\'e}e} 2014a]{eg99}. In addition, the southern and
northern tangencies for the Near 3 kpc Arm\footnote{The identification
  of the ``3 kpc Arm'' dates back to 1960s
  \citep[e.g.,][]{vro57,okw58}. There still are some debates on its nature
  \citep[][]{gcm+11}, which could be an expanding ring-like structure
  \citep[e.g.,][]{van71,cd76,sev99}, non-expanding resonance feature,
  or elliptical streamlines \citep[e.g.,][]{pet75}, or spiral arms
  \citep[e.g.,][]{fu99,beg03}. Two tangencies are observed for this
  structure in the inner Galaxy regardless of its nature. In this
  work, we take the name ``3 kpc Arm'' when referring to the feature.}
have also been reported \citep[e.g.,][]{cd76,ban80,dt08}. The
directions of arm tangencies are regarded as important observational
constraint on the Galaxy spiral arms
\citep[e.g.,][]{bur71,bur73,gg76,eg99,ds01,ne2001,rus03,cbm+09,hhs09,hh14,val15}. A
detailed literature survey for spiral arm tangency directions can be
found in \citet{val14}.

The definitions of an arm tangency are often slightly different in
literature, which makes direct comparison and even statistics
difficult. For example, the arm tangency in \citet[][see their
  Sect. 4.6]{eg99} is determined as the outer edge of a spiral arm
where the velocity jump occurs. The arm tangency can be traced by
discontinuities in the integrated CO emission, as mentioned by
e.g. \citet{bcmn00}, \citet{gcb+87}, and \citet{amb90}, or by the
solid-body like kinematics in the rotation curve derived from CO
survey \citep[e.g.,][]{lbcm06}. Spiral arms produce the deviations
from the circular velocity, so that the bumps in the rotation curves
are anomalous velocities associated with streaming motion \citep[see
  e.g.][]{bs70,bur73,md07}. The arm tangency can also be identified
from peak features \citep{dri00,bact89} or local maxima \citep{swh10}
in the longitude plot for integrated infrared emission or line
intensities. \citet{cfg+11} identified the Norma tangency as a dense
concentration of methanol masers. The arm tangency can also be found
by fitting the distributions of spiral tracers in the Galactic plane
or by fitting the features in the longitude-velocity diagram with a
spiral arm model \citep[e.g.,][]{hh14}. The derived arm tangency may
be influenced by the position uncertainties of tracers. The derived
directions of arm tangencies by different authors could be different
as large as the half of arm width even from the same observational
data.

\begin{figure*}
\centering\includegraphics[width=0.95\textwidth]{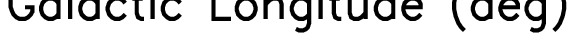}
\caption{Surface density of old stars as a function of the Galactic
  longitude deduced from the survey data of GLIMPS 4.5 $\mu$m (black),
  2MASS {\it K$_s$} band (red), 2MASS {\it H} band (green) is shown in
  {\it panel} $<$1$>$.
The integrated intensity of the 1.4~GHz RRLs over the velocity
  range of $\pm$15~km~s$^{-1}$ from the tangency point velocity of a given
  direction is plotted against the Galactic longitude in {\it panel}
  $<$2$>$.
The surface densities of HII regions (black) and 6.7-GHz methanol
masers (red) are shown in {\it panel} $<$3$>$.
 The integrated intensity for molecular gas $^{12}$CO(1$-$0) (black)
 and $^{13}$CO(1$-$0) (red) and neutral hydrogen HI, again over
 $\pm$15~km~s$^{-1}$, are plotted in {\it panel} $<$4$>$ and {\it panel}
 $<$6$>$, respectively.
The surface density for dense cores of molecular clouds detected from
the ATLASGAL (black) and the HOPS (red) are plotted in {\it panel}
$<$5$>$.
 Peaks for spiral arm tangency directions are indicated by arrows with
 different colors. The complex features in the regions close to the
 Galactic center direction ($20^\circ > l > 340^\circ$) are not the
 scope of this article and plotted as gray lines. See text for
 references for the survey data, and for discussions for the
 identifications of spiral arm tangencies.}
\label{all_plot}
\end{figure*}

There are some potential pitfalls on identifications of arm tangencies
from observations. For example, velocity crowding, including the
streaming motions \citep[e.g.,][]{bur73}, and concentrations of
individual clouds can result in the ``bump'' features in the
longitude-velocity map or the number counts of objects. The
  clumps or cores of nearby clouds or star forming regions could be
mistaken as the longitude concentrations of farther objects and
then be misinterprested as arm tangencies. In addition, gas
observations for tangencies could be complicated by optical depth
effects for some lines, such as the known self-absorption effect for
the HI 21 cm line and CO line. Observations of small distant clouds
can suffer beam dilution effect. To identify arm tangencies properly,
the multi-wavelength survey data for different Galactic components
should be considered together, because the problems discussed above
may present in one or two datasets, but not in all datasets.

The spiral arms traced by old stars and molecular gas in our Milky Way
are believed to be spatially coincident \citep[e.g.,][]{val14}.  A
spatial separation ($\sim$ 100 pc $-$ 300 pc) between different arm
tracers, e.g., $^{12}$CO, $^{13}$CO, HI gas, cold dust, methanol
masers, hot dust, was recently suggested by \citet{val14a}, based on
the statistics of arm tangency direction values in literature without
considering possible different definitions. To show the possible
offsets between spiral arms traced by different Galactic components,
i.e., old stars, ionized gas, molecular gas, atomic gas, we reassess
the spiral arm tangencies by using survey data in literature for
different Galactic components.


\section{Survey data for different Galactic components}

\subsection{Stellar component}
Since the spiral structure traced by the stellar component were
obtained by using the COBE survey \citep[][]{dri00,ds01}, data quality
of two following surveys\footnote{In fact, the Wide-field Infrared
  Survey Explorer \citep[WISE,][]{wise} also finished an all sky
  survey at 3.4 $\mu$m, 4.6 $\mu$m, 12 $\mu$m, and 22 $\mu$m. The WISE
  data have a Point Spread Function of
    6$^{\prime\prime}-12^{\prime\prime}$ in the 3 $\mu$m to 24 $\mu$m
    bands, and suffer from confusion near the Galactic midplane,
  hence, are not used in this work for distributions of old
  stars. While HII regions used in this paper were identified from
  WISE data \citep{abb+14}, see Sect. 2.2.} has been improved
significantly in the resolution and sensitivity with a greatly reduced
extinction:

1) Two Micron All Sky Survey~\citep[2MASS,][]{2mass} observed almost
the entire celestial sphere in three near-infrared bands, i.e., {\it
  J} (1.25~$\mu$m), {\it H} (1.65~$\mu$m) and {\it K$_s$}
(2.16~$\mu$m), produced a point source catalog containing 470, 992,
970 sources. The sensitivity (S/N = 10) is 15.8~mag, 15.1~mag, and
14.3~mag for the {\it J}, {\it H}, and {\it K$_s$} bands,
respectively.

2) {\it Spitzer}/GLIMPSE (Galactic Legacy Infrared Midplane
Extraordinaire) is a survey of the inner Galactic plane with the {\it
  Spitzer} space telescope in four mid-infrared bands, i.e.,
3.6~$\mu$m, 4.5~$\mu$m, 5.8~$\mu$m and 8~$\mu$m \citep{glimpse},
providing a 3$\sigma$ sensitivity for point sources as being 15.5~mag,
15.0~mag, 13.0~mag and 13.0~mag, respectively \citep{cbm+09}. The
GLIMPSE has a comparable resolution of $\sim 1"$ and a sensitivity of
$\sim$ 0.5~mJy to the 2MASS but with a greatly reduced extinction
\citep[][]{ben08}. We note that about 90\% of all the GLIMPSE
stars are red giants, and a good fraction of them appear to be the
red-clump giants \citep[][]{cbm+09} which have a very narrow
luminosity function around an absolute magnitude $M_K= -
1.61\pm0.03$~mag with an Gaussian width of $\sigma_K =
0.22\pm0.03$~mag \citep[e.g.,][]{alv00,chg+07}.

The point sources detected by GLIMPSE and 2MASS within the Galactic
latitude range of $|b|<1^\circ$ and the magnitude range of $\Delta m =
6.5-12.5$ are used in this paper for determination of spiral arm
tangencies traced by old stars\footnote{The GLIMPSE Point Source
  Catalog is truncated at 6.5~mag due to the detector nonlinearity for
  bright sources \citep[e.g., see Fig.3. of][]{bcb05}. In the inner
  Galaxy ($|l|< \sim$ 40$^\circ$), the GLIMPSE is confusion-limited at
  13.3$-$13.6~mag \citep[][]{cbm+09}. Following \citet{bcb05} and
  \citet{cbm+09}, we adopt 12.5~mag as the fainter limit for the
  GLIMPSE and 2MASS source plots in Fig~\ref{all_plot}. The influence
  of the magnitude range on the derived spiral arm tangencies is
  tested and discussed in Sect.3.1.}.  Surface number density of stars
(per deg$^{2}$) is plotted against the Galactic longitude in the top
panel of Fig.~\ref{all_plot}. We ignore the regions close to the
Galactic center direction (e.g., $20^\circ > l > 340^\circ$) where the
bump features may be the superposition of structure features in the
Galactic bar(s), bulge, and spiral arms. We focus only on the
longitude regions of $ 20^\circ < l < 340^\circ$, where our plots are
consistent with the results given by \citet[][see their
  Fig.~14]{cbm+09}, \citet[][see their Fig.~2]{ben09} and also
\citet[][see their Fig.~1]{dri00}. Four broad and prominent bumps can
be identified, which are interpreted in this paper as indication of
spiral arm tangencies for the Scutum Arm ($l\sim 32^\circ$), the
Centaurus Arm ($l\sim 308^\circ$) and the northern and southern ends
of the Near 3 kpc Arm ($l\sim 27^\circ$ and $l\sim 338^\circ$) though
the stellar bump near $l\sim27^\circ$ can also be interpreted by an
in-plane bar or a ring \citep[see Sect.1 of][]{lhg01}. The peak of the
bumps are related to the density maximum in stellar spiral arms (see
Sect. 3.2).

\subsection{Ionized gas and star formation sites}

Ionized gas exists in the interstellar medium in general in three
forms: individual HII regions, diffuse warm ionized gas, and hot
ionized gas \citep[][]{leq05,fer01}. Radio recombination lines are
best tracers of ionized gas \citep[][]{gs09}. However, the global
properties of ionized gas in the Milky Way revealed by RRLs are not
well explored yet \citep[e.g.,][]{tbd+14,lmt+13,acd+14,bbo+15}. At
present, the HIPASS survey of the Galactic plane
\citep[][]{add+10,acd+14} has the largest sky coverage for mapping the
RRLs in the region of $l = 196^\circ - 0^\circ - 52^\circ$ and $|b|
\leq 5^\circ$ at 1.4~GHz with a resolution of 14.4~arcmin, picking up
the RRLs of H168$\alpha$, H167$\alpha$ and H166$\alpha$ with a rms
noise per channel of about 2.8~mK.

In the panel $<$2$>$ of Fig.~\ref{all_plot}, the RRL line intensity
integrated over $|b|<$ 2$^\circ$ and a velocity range of $\Delta V=
[V_t~-$ 15 km~s$^{-1}$, $V_t~+$ 15 km~s$^{-1}]$ around the terminal
velocity $V_t$ is plotted against the Galactic longitude. The velocity
range $\Delta V$ is so chosen to include the gas around the tangencies
but exclude the foreground and background RRL emission. The terminal
velocity $V_t$ is calculated with a flat rotation curve and the IAU
standard circular orbital speed at the Sun, $\Theta_0$ = 220
km~s$^{-1}$. The influences of the adopted rotation curves,
$\Theta_0$, and velocity range $\Delta V$ will be tested and discussed
in Sect. 3.1.

In the RRL plot, each spiral arm shows a corresponding distinct
bump. The bumps indicate that ionized gas traced by H168$\alpha$,
H167$\alpha$ and H166$\alpha$ are concentrated to spiral arms. The
bump peaks indicate the density maximums of ionized gas in spiral arms
(see Sect. 3.2) and hence are adopted as the ``observed'' arm
tangencies of ionized gas. Some weaker bumps in the plots may indicate
some individual clouds or arm spurs/branches in the interarm
regions. The significance of a bump is evaluated by the ratio of the
bump peak value to the fluctuation $\sigma$ which is estimated from
the fitting residual outside the range of a tangency.

The Centaurus arm is intriguing to have two distinct RRL bumps near
$l\sim311^\circ$ and $l\sim305^\circ$, which are also shown in the CO
and HI plots and coincident with the dips in the rotation curve of
\citet[][see their Fig. 8]{md07}.
  
HII regions are the zones of ionized gas surrounding young massive
stars or star clusters. They are primary tracer of spiral arms. HII
regions can be detected from the whole Galactic disk without
extinction, even as far as $\sim$20 kpc from the Sun
\citep[see][]{hh14}. The largest catalog of Galactic HII regions or
candidates to date is recently given by \citet{abb+14}, who identified
HII regions according to the mid-infrared morphology in the WISE
survey data. More than 8400 HII regions and candidates are identified,
about 1500 of them were previously known. However, most of the HII
regions do not have line velocity information.

The 6.7 GHz methanol masers are good tracers of the early evolutionary
stage of massive stars. About 1000 Galactic methanol masers have been
detected
\citep[e.g.,][]{pmb05,elli07,xlh+08,cbhc09,gcf+09,gcf+10,gcf+12,cfg+10,cfg+11,fcf10,oah+13,sxc+14,bfc+15}. The
catalog of 6.7-GHz methanol masers collected by \citet[][]{hh14} from
literature is used in this work.

The number density of HII regions and methanol masers is plotted
against the Galactic longitude in the panel $<$3$>$ of
Fig.~\ref{all_plot}. However, the identification of spiral arm
tangencies in the plot is not straightforward as the bump features are
not so clear. The star formation sites are very clumpy in general, and
some bumps may be related to arm spurs or arm branches or clumpies in
some nearby individual clouds, not to the major gaseous spiral
arms. In addition, no velocity constraints were made in the plot due
to the lack of velocity information for most HII regions. Some bump
features are probably related to the star formation sites in the
Sagittarius-Carina Arm and/or Scutum-Centaurus Arm. It is clear that
spiral arm tangencies can be well-identified by RRL, CO and HI
velocity integrated plots, and then we check the coincident features
in the number density plot for HII regions and methanol masers. The
bumps near the tangency directions can be found for the northern part
of the Near 3 kpc Arm ($l\sim24^\circ$), the Scutum Arm ($l\sim
32^\circ$), the Sagittarius Arm ($l\sim 50^\circ$), the Centaurus Arm
($l\sim 306^\circ$, 312$^\circ$), the Norma Arm ($l\sim 328^\circ$),
and the southern part of the Near 3 kpc Arm ($l\sim 337^\circ$), which
we marked arrows in Fig~.\ref{all_plot}. These bump peaks indicate the
dense concentration of HII regions and methanol masers in spiral arms,
hence are taken as the ``observed'' arm tangencies for star formation
sites. The density excesses near the northern tangency of the Near 3
kpc Arm ($l\sim 24^\circ$) and near the Carina tangency ($l\sim$
282$^\circ$) are not prominent with a significance of only
$0.6\sigma-2.6\sigma$.
%
%

\subsection{Molecular gas and high density sites}

Carbon monoxide (CO) is a tracer for molecular gas in galaxies.  By
far the most widely-used survey of Galactic $^{12}$CO(1$-$0) was given
by \citet{dht01}, which covers the entire Galactic plane and extends
at least six degrees in the Galactic latitude. The angular resolution
is 8.4$^\prime$ $-$ 8.8$^\prime$. The rms noise per channel is about
0.3 K. The velocity coverage is from $-$260 km~s$^{-1}$ to $+$280
km~s$^{-1}$ in the frame of the Local Standard of Rest (LSR).

Emission of $^{13}$CO(1$-$0) suffers less extinction than that of
$^{12}$CO(1$-$0), hence it is a better tracer of column density for
the molecular gas. The Galactic Ring Survey \citep[GRS,][]{grs} of
$^{13}$CO(1$-$0) covers the longitude range from 18$^\circ$ to
55.7$^\circ$ and the latitude range of $|$b$|$ $\leq$ 1$^\circ$. The
angular resolution is about 46$^{\prime\prime}$ and the sampling is
22$^{\prime\prime}$. The sensitivity is about 0.4 K. The LSR velocity
coverage is from $-$5 km~s$^{-1}$ to $+$135 km~s$^{-1}$.

In the panel $<$4$>$ of Fig.~\ref{all_plot}, the variation of CO
intensity integrated over $|b|<$ 2$^\circ$ and a velocity range of
$\Delta V= [V_t~-$ 15 km~s$^{-1}$, $V_t~+$ 15 km~s$^{-1}]$ around the
terminal velocity is plotted against the Galactic longitude
\citep[e.g., see also][]{gcb+87}. We see that near every spiral arm
tangency there is a corresponding bump in the CO plot. The bump peaks
in the CO plot is related to the density maximums of molecular gas in
spiral arms (see Sect. 3.2). Therefore molecular clouds are good
tracers of spiral structure \citep[e.g.,][]{hh14}, and the spiral arm
tangencies can be shown by molecular gas. We note that around the
tangency direction of the Centaurus arm, there are two distinct peaks
of CO intensity, stronger one near $l\sim312^\circ$ and a weaker one
near $l\sim306^\circ$.

High density regions of molecular gas comprise only $\sim$ 7\% of the
mass of molecular clouds \citep[][]{bh14}, but they are the birth
place of massive stars or star clusters. The best way to unambiguously
search for high density regions is to survey the optically thin
emission of dust in the millimeter/submillimeter regime
\citep[][]{cus+14,awb00}. There have been some impressive progress in
recent years, e.g., the Bolocam Galactic Plane Survey
\citep[][]{rdg+10,agd+11,ggr+13}, the APEX Telescope Large Area Survey
of the Galaxy \citep[ATLASGAL,][]{csu+13,cus+14}, and the {\it
  Herschel Space Observatory} survey \citep[Hi-Gal,][]{higalb,higala}.
The ATLASGAL at 870 $\mu$m has been finished and cover a large portion
of the inner Galactic plane ($-$60$^\circ$ $\leq l \leq$ +60$^\circ$,
and $-$1.5$^\circ$ $\leq b \leq$ +1.5$^\circ$; $-$80$^\circ$ $\leq l
\leq$ $-$60$^\circ$, and $-$2.0$^\circ$ $\leq b \leq$
+1.0$^\circ$). More than 10000 compact sub-millimeter sources (dense
clumps) were identified \citep{csu+13,cus+14}.

In addition, the dense regions of molecular gas can be detected by
using molecular lines tracing high density environments, e.g.,
HCO$^+$, NH$_3$ \citep[][]{hops12,ses+13}.
The H$_2$O Southern Galactic Plane Survey (HOPS) maps the inner
Galactic plane ($-$70$^\circ$ $> l >$ 30$^\circ$, $|b|<$ 0.5$^\circ$)
using the Mopra 22m telescope at 12 mm wavelengths \citep[19.5$-$27.5
  GHz,][]{hops}, and detect 669 dense clouds by using NH$_3$(1,1)
transition \citep{hops12}.

The number density of the dense clumps from the ATLASGAL and HOPS is
plotted against the Galactic longitude in the panel $<$5$>$ of
Fig.~\ref{all_plot}. The arm tangencies for the northern part of the
Near 3 kpc Arm ($l\sim$ 24$^\circ$), the Scutum Arm ($l\sim$
32$^\circ$), and the Sagittarius Arm ($l\sim$ 50$^\circ$) can be
easily identified from the longitude plot without ambiguity. Similar
to the number density plots for HII regions and methanol masers, the
bump features in the fourth Galactic quadrant are complex, and need to
be compared to arm tangency directions obtained from the RRL/CO/HI
plots. The identified bump peaks indicate the concentration of dense
clumps around spiral arms, which can be taken as the ``observed'' arm
tangencies from high density gas clumps.

\begin{figure}
\centering\includegraphics[width=0.45\textwidth]{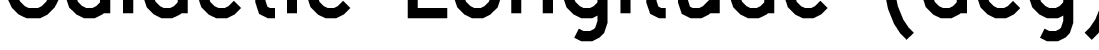}
\caption{An example of fitting the longitude plot of old stars for the
  tangency of the Scutum Arm and the northern tangency of the Near 3
  kpc Arm. The surface density of stars are deduced from the GLIMPSE
  4.5 $\mu$m data. The ``baseline'' (green) is subtracted first (black
  crosses in the bottom), and then three Gaussians (yellow) are fitted
  to the data.}
\label{spitzer}
\end{figure}

\setlength{\tabcolsep}{0.8mm}
\begin{table*}
  \caption{The Galactic longitudes of ``observed'' spiral arm
    tangencies of the Milky Way recognized from the survey data of old
    stars, RRLs, HII regions, methanol masers, CO, dense clumps, and
    HI. The measured bump widths in FWHM are given in brackets.  The
    median values of arm tangencies for stars (GLIMPSE/2MASS) and gas
    (RRLs, HII regions, masers, CO, dense clumps, and HI) are given in
    the bottom. Values with ``$\ast$'' have the significance of
    identified bump less than $3\sigma$.}
\begin{center}
\begin{tabular}{lccccccc}
  \hline
  \hline
Survey Data   & Near 3 kpc North & Scutum & Sagittarius & Carina & Centaurus & Norma & Near 3 kpc South\\
  & $(^{\circ})$  & $(^{\circ})$ & $(^{\circ})$ & $(^{\circ})$ & $(^{\circ})$& $(^{\circ})$ & $(^{\circ})$   \\
  \hline
GLIMPSE 3.6 $\mu$m          & 26.9 $[3.2]$   & 32.6 $[1.7]$  &--     &--   & 307.5 $[4.0]$   &-- &  338.3 $[4.8]$      \\
GLIMPSE 4.5 $\mu$m          & 26.9 $[3.8]$   & 32.6 $[1.8]$  &--     &--   & 307.6 $[5.1]$   &-- &  338.3 $[5.1]$      \\
GLIMPSE 5.8 $\mu$m          & 27.1 $[3.3]$   & 32.7 $[1.6]$  &--     &--   & 307.4 $[4.3]$   &-- &  337.7 $[2.4]$      \\
GLIMPSE 8.0 $\mu$m          & 27.4 $[1.5]$   & 32.8 $[1.2]$  &--     &--   & 307.3 $[1.9]$   &-- &  337.8 $[0.9]$      \\
2MASS $K_s$ (2.16 $\mu$m)   & 27.0 $[2.3]$   & 32.6 $[1.5]$  & 55.0$^\ast$ $[2.7]$  & -- & 307.5 $[2.9]$  & -- & 338.3 $[3.6]$       \\
2MASS $H$ (1.65 $\mu$m)     & 27.0 $[2.1]$   & 32.6 $[1.4]$  & 55.0$^\ast$ $[2.6]$  & -- & 307.5 $[2.5]$  & -- & 338.3 $[2.9]$      \\
2MASS $J$ (1.25 $\mu$m)     & 27.1 $[2.1]$   & 32.5 $[1.4]$  & 55.0$^\ast$ $[2.3]$  & -- & 307.1 $[2.1]$  & -- & 338.3 $[1.7]$      \\
 &  &  &  &  &   & &       \\
1.4-GHz RRLs     & 24.6 $[0.7]$  & 30.8 $[0.7]$  & 49.2 $[0.8]$ & 284.3 $[0.6]$  & 305.4 $[0.7]$, 311.2 $[0.9]$  & 329.3 $[1.9]$  & 336.9$^\ast$ $[0.6]$ \\
WISE HII regions & 23.5$^\ast$ $[2.1]$ & 30.6 $[1.5]$  & 49.4 $[1.0]$ & 283.3$^\ast$ $[2.1]$ & 305.5 $[0.8]$, 311.7 $[1.8]$  & 328.1 $[2.2]$  & 337.2 $[1.0]$           \\
6.7-GHz methanol masers & 24.6$^\ast$ $[0.7]$  & 30.8 $[0.9]$  & 49.3 $[0.8]$ & 284.5$^\ast$ $[0.5]$ & 305.5 $[0.9]$, 312.2 $[1.1]$   & 329.3 $[2.9]$   & 337.0 $[2.8]$ \\
$^{12}$CO (DHT2001)      & 24.4 $[0.3]$  & 30.5 $[2.0]$  & 49.4 $[0.5]$  & 282.0 $[2.7]$  & 305.7 $[5.2]$, 311.0 $[4.6]$  & 328.3 $[2.2]$ & 336.7$^\ast$ $[0.8]$        \\
$^{13}$CO (GRS)          & 24.4 $[0.4]$  & 30.5 $[1.6]$  & 49.4 $[0.5]$  & --       & --             & --     & --     \\
ATLASGAL dust sources   & 23.8 $[1.7]$  & 30.7 $[0.9]$  & 49.2 $[0.7]$  & 284.2$^\ast$ $[0.5]$  & 305.3 $[0.8]$, 311.7$^\ast$ $[1.0]$  & 327.2 $[2.5]$  & 337.5 $[2.6]$ \\
HOPS NH$_3$ sources     & 23.7 $[2.3]$   & --   &   --     & --       & 305.8 $[1.5]$, 309.1 $[0.6]$   & 327.8 $[2.1]$  & 338.4 $[2.5]$        \\
HI (LAB)                & --   & 30.8 $[2.3]$  & 50.8 $[5.8]$  & 283.0 $[9.3]$    & 304.3 $[14.1]$, 310.4 $[6.4]$   & 328.0 $[4.5]$ & 336.8 $[5.7]$        \\
\hline
Median for old stars         & 27.0   &  32.6  & 55.0  &  --   & 307.5   & --    & 338.3  \\
Median for gas               & 24.4   &  30.7  & 49.4  &283.8  & 305.5, 311.2   & 328.1 & 337.0  \\
  \hline\hline
\end{tabular}
\end{center}
\label{tan}
\end{table*}

\begin{figure}
\centering\includegraphics[width=0.45\textwidth]{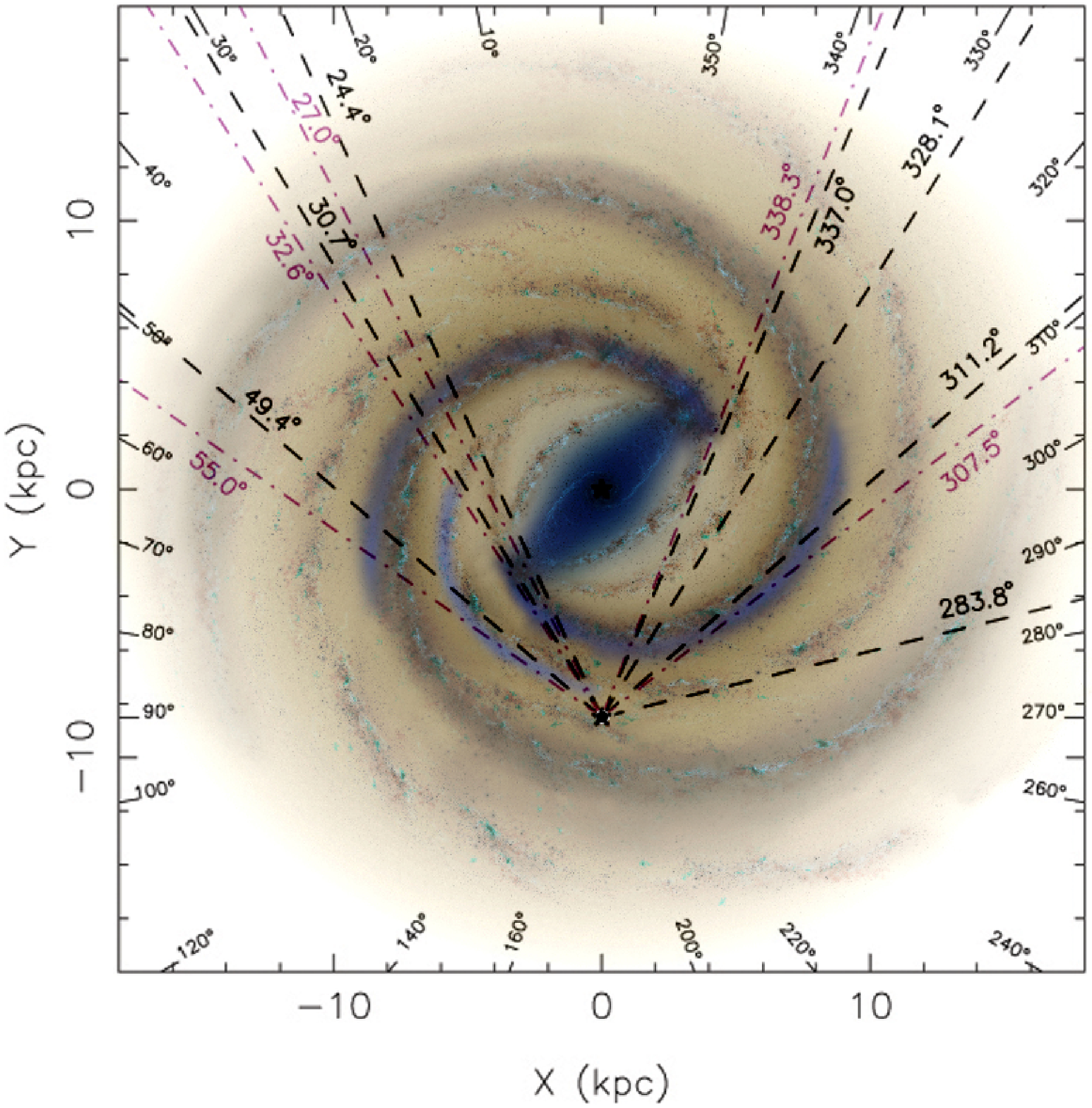}
\caption{The longitude values of arm tangencies for different spiral
  components comparing with those determined by $^{12}$CO in this
  paper. Filled symbols indicate the assessments in this paper, and
  the open symbols stand for the values in literature
  \citep[see][]{eg99,hh14,val14}.}
\label{com}
\end{figure}

\subsection{Atomic gas: HI}

Atomic gas in our Galaxy has been extensively observed by HI 21cm line
with low resolution in early days and more recently at
high resolution \citep[][]{sgps,vgps}. The Leiden/Argentine/Bonn (LAB)
survey covering the entire sky is the most sensitive HI survey to date
\citep{lab_hi}, which merges the Leiden/Dwingeloo survey of the
northern sky \citep{lab_a} and the Instituto Argentino de
Radioastronomia Survey of the southern sky \citep{lab_c,lab_b}. The
angular resolution is $\sim$ 36$^\prime$. The LSR velocity coverage is
from $-$450 km~s$^{-1}$ to $+$400 km~s$^{-1}$. The rms noise of the
data is 70 mk $-$ 90 mk.

Similar to the RRL plot and the CO plot, the integrated HI intensity
over $|b|<$ 2$^\circ$ and a velocity range of $\Delta V= [V_t~-$ 15
  km~s$^{-1}$, $V_t~+$ 15 km~s$^{-1}]$ around the terminal velocity is
plotted in the panel $<$6$>$ of Fig.~\ref{all_plot}. The atomic gas is
not obviously concentrated to spiral arms as molecular gas \citep[also
  see][]{ns03,ns06}, so that the bump features in the HI plot are not
as prominent as those in the RRL and CO plots. The known arm
tangencies, however, can be recognized except for the northern
tangency of the Near 3 kpc Arm ($l\sim24^\circ$). The bump peaks in
the HI plot indicate the density maximums of atomic gas in spiral
arms.

\section{Arm tangencies for different components}

\subsection{Measuring arm tangencies from the longitude plots}

As discussed above, the peaks of the bumps in the longitude plots
correspond to the arm tangencies (see Fig.~\ref{all_plot}).  To
measure the peak position, we first fit the ``baseline'' with a
second-order polynomial function outside the tangency range which
should represent the general contribution from the Galaxy disk. The
baseline is then subtracted from data to show bumps more
clearly. After that a single Gaussian or multi-gaussians are fitted to
a bump, as shown in Fig.~\ref{spitzer}. The peak position of the
fitted Gaussian(s) is adopted as the longitude direction of
``observed'' arm tangency, and the bump width is derived as well to be
the full width at half maximum (FWHM), as listed in
Table~\ref{tan}. We noticed that for each arm the tangency longitudes
derived from the gaseous components (RRLs, star formation sites, CO,
HI, dense clumps) have similar values, with a much smaller difference
than various values in literature (see Fig.~\ref{com}). We verified
that results change less than 1$^\circ$ if other fitting functions are
used, e.g., a single Gaussian plus a first-order polynomial function.
 
Note also that a flat rotation curve with the IAU standard circular
orbital speed at the Sun of $\Theta_0$ = 220 km~s$^{-1}$ is adopted to
calculate the terminal velocity $V_t$, and then the velocity range of
$\Delta V= V_t~\pm 15$~km~s$^{-1}$ for the integrated line intensity
is used to derive the longitude plots for RRLs, CO and HI in
Fig.~\ref{all_plot}. We tested and found that if the other rotation
curves, e.g. the one given by \citet{bb93}, \citet{cle85}, and
\citet{fbs89}, or different $\Theta_0$ \citep[238 $-$ 240
  km~s$^{-1}$,][]{sch12,rmb+14}, or even the different velocity ranges
such as $\pm 5- \pm 25$~km~s$^{-1}$, are used, the measured arm
tangencies change less than $\sim$0.5$^\circ$ and the measured bump
widths change less than $\sim$1$^\circ$. So does the magnitude range
for the old stars.

\begin{figure}
\centering\includegraphics[width=0.47\textwidth]{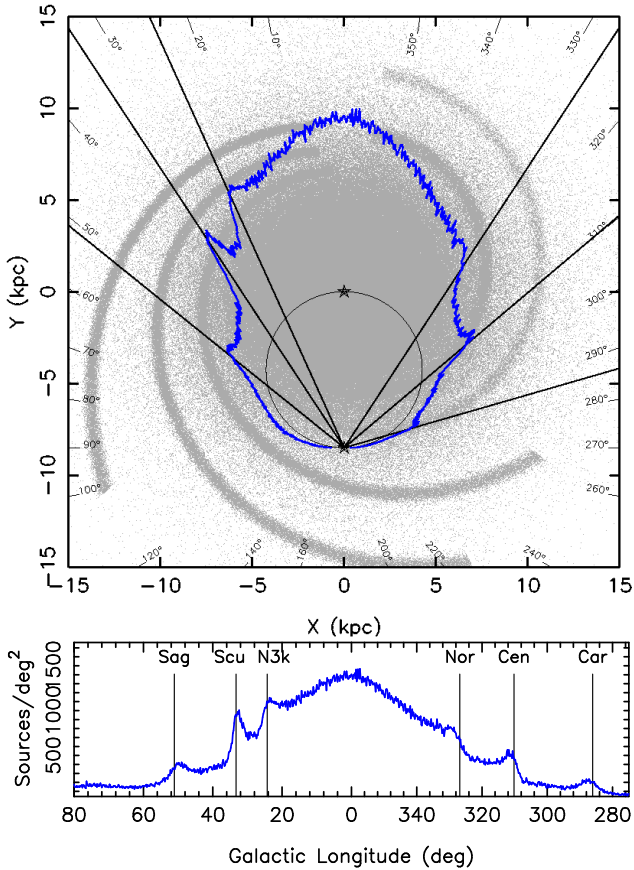}\\
\centering\includegraphics[width=0.47\textwidth]{simu_lv.ps}\\
\caption{The density model ({\it the upper panel}) and the surface
  density of old stars as a function of the Galactic longitude ({\it
    the lower panel}). The ``observed'' arm tangencies can be
  identified by the bump peaks. The ``true'' density peaks near spiral
  arm tangencies in the model are indicated by solid lines, which are
  displaced from the ``observed'' bump peaks.}
\label{simu}
\end{figure}

\begin{figure}
\centering\includegraphics[width=0.45\textwidth]{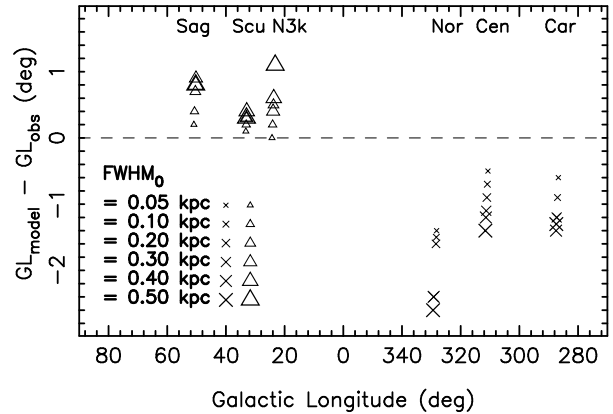}
\caption{The ``observed'' longitude values for arm tangencies
  ($GL_{obs}$) deduced from various simulated models with a different
  input arm width (FWHM$_{\rm 0}$). The ``observed'' tangency
  directions are shifted to the interior to the ``true'' arm density
  peaks of the model ($GL_{model}$).}
\label{sobs}
\end{figure}

\subsection{``Observed'' tangencies and the true density peaks}

To explain the detailed features in the plots of Fig.~\ref{all_plot},
an illuminating model should include disk, bulge, Galactic bar(s),
spiral arms, and also interstellar extinction and luminosity function
\citep[e.g.,][]{ol93,dri00,rcb+12,pjl13}. Here, we try to understand
the interplay between the observed bumps for arm tangencies and the
true density peaks in spiral arms. For this purpose, we make a
simplified model as shown in Fig.~\ref{simu}, which consists of two
density components \citep[see ][]{ds01}:
\begin{equation}
\rho = \rho_{disk} + \rho_{arm} = A_{disk} exp(-r/h_r) +
A_{arm}\sum_iexp(-d_i/w_i)^2.
\label{eq0}
\end{equation}
Here, $\rho_{disk}$ is an axisymmetric exponential disk, $\rho_{arm}$
comes from spiral arms, $r$ is the Galactocentric distance, $h_r$ is
the scale length, $d_i$ is the distance to the nearest spiral arm,
$w_i(r) \propto r$ indicates the half-width of the $i$th arm, and
$A_{disk}$ and $A_{arm}$ are the density normalization for the disk
and spiral arms, respectively. The density profile acrosses an arm is
taken as Gaussian.  The polynomial-logarithmic model derived by
fitting to the distribution of HII regions by \citet[][see the {\it
    left} panel in their Fig. 11]{hh14} is adopted for the spiral
arms, which have the known density peaks at arm tangencies of spiral
arms. We put 10$^6$ particles randomly to match the density model in
Fig.~\ref{simu}.

According to \citet{bm98}, the number of stars of type $s$ with
apparent magnitude between $m_1$ and $m_2$ in a solid angle $d\Omega$
in the ($l,b$) direction is:
\begin{equation}
N_s(m_1,m_2,l,b)d\Omega = \int^{m_2}_{m_1}dm \int^{\infty}_0
\rho_s(R,l,b,M) \Phi_s(M)R^2 dR d\Omega,
\label{eq1}
\end{equation}
here, $R$ is the heliocentric distance, $M$ is the absolute magnitude,
$\rho_s$ is the stellar density near the Galactic plane, and $\Phi_s$
is the luminosity function of stars of type $s$. Because a good
fraction of GLIMPSE detected stars appear to be red-clump giants
\citep[][]{cbm+09}, it would be reasonable to assume that the bump
features shown in the infrared star counts are dominated by red-clump
giants. We assume that the stellar density is proportional to the
density model given in Eq.~(\ref{eq0}). Considering that the GLIMPSE
mid-infrared survey allowed for the nearly-extinction free detection
of stars in the inner Galactic plane \citep[][]{ben08}, and also for
simplicity, we neglect the extinction effect. The luminosity function
of red-clump giants given by \citet[][see their Fig.3]{alv00} is
normalized and adopted, and the apparent magnitude range between
$m_1=6.5$ and $m_2=12.5$ is adopted in the calculations. Then the
longitude plot for the surface density of stars can be derived as
shown in the lower panel of Fig.~\ref{simu}, with clear bump features
for arm tangencies.

We find that the directions for ``observed'' arm tangencies deviate
from the longitudes of the ``true'' density maximums of spiral arms
near the tangencies by shifting to the inner side of about 1$^\circ$
in general (see Fig.~\ref{sobs}), depending on the input arm width in
the model. Near the Norma tangency, the shift could be as large as
2.5$^\circ$ if the arm has a width (FWHM) of 0.4 kpc $-$ 0.5 kpc.

\begin{figure*}
\centering\includegraphics[width=0.49\textwidth]{scutumV3.ps}
\centering\includegraphics[width=0.49\textwidth]{centaurusV3.ps}
\caption{The longitude plots after ``baseline'' subtracted (see
  Fig.~\ref{all_plot} and Fig.~\ref{spitzer}) for the tangency of the
  Scutum Arm and the northern tangency of the Near 3 kpc Arm ({\it
    left panel}) and the Centaurus Arm ({\it right panel})
, normalized by the maximum values in each plot. The tangencies are
  indicated by arrows together with typical uncertainties less than $
  1^\circ$.}
\label{scutum}
\end{figure*}

Similar simulations for gas distributions can be obtained to get the
modeled longitude plots of integrated intensities. Take the CO data as
an example. The integrated emission $W(l,b)_{\rm CO}d\Omega = \int
T_R(l,b,v) dvd\Omega$ is commonly assumed to be proportional to the
column density of molecular hydrogen $N(H_2)$, e.g., $N(H_2)=X
W(CO)$. Here $X$ is the conversion factor, $T_R(l,b,v)$ is the CO line
temperature \citep[e.g.,][]{gcb+87}. Hence, $dN(H_2) = \rho(R,l,b)dR =
XT_R(l,b,v)dv$, where $\rho(R,l,b)$ is the gas density, $R$ is the
heliocentric distance. Then the integrated intensity of CO:
\begin{equation}
\int T_R dvd\Omega \propto \int \rho(R,l,b)dRd\Omega
\label{eq2}
\end{equation}
can be obtained from the density model (see Fig.~\ref{simu}). We get
the same conclusion for gas arms as that for the stellar arms.

More sophisticated model could be constructed to explain the real
data.  For example, the infrared stars in the GLIMPSE and 2MASS
surveys consist of a mixture of different types of stars with
different luminosity functions, and distant stars should suffer from
extinction effect. The optical depth and beam dilution effects should
be considered in the gas density model. However, the very simplified
models discussed above seem to be good enough to illustrate the
interplay between the observed bumps and the true density peaks near
the spiral arm tangencies.

\subsection{Displacements of spiral arms outlined by different tracers}

The measured directions for tangencies are not exactly the real
directions of density peaks of spiral arms near the tangencies, with a
slight shift to the interior to the arm. However such a shift happens
to all kinds of spiral tracers, various gas arms or stellar
arms. Therefore, the {\it relative} longitude positions of arm
tangencies shown by different components can be used to check the
spatial coincidence of different spiral arm components.

For the tangencies of the Scutum-Centaurus Arm and the northern and
southern parts of the Near 3 kpc Arm, the consistent longitude is
found for the stellar arm tangency from the infrared data of GLIMPSE
and 2MASS (see Fig.~\ref{scutum} and also Table~\ref{tan}) although
the extinction is significantly less at GLIMPSE 4.5 $\mu$m band than
the 2MASS $J$, $H$, and $K_s$ bands (Fig.~\ref{all_plot}).

Arm tangencies shown by RRLs, HII regions, methanol masers, CO, dense
clumps, and HI data have more or less similar longitudes (see
Fig.~\ref{scutum}) with a small difference of $\leq 1.1^\circ$ ($\sim$
150 pc) for the northern tangency of the Near 3 kpc Arm, $\leq
0.3^\circ$ ($\sim$ 40 pc) for the Scutum tangency, $\leq 1.6^\circ$
($\sim$ 150 pc) for the Sagittarius tangency, $\leq 2.5^\circ$ ($\sim$
90 pc) for the Carina tangency, $\leq 2.1^\circ$ ($\sim$ 260 pc) for
the Norma tangency, and $\leq 1.7^\circ$ ($\sim$ 230 pc) for the
Southern Near 3 kpc tangency, which are comparable to the typical
uncertainty of the measured arm tangencies ($\sim 1^\circ$), and
smaller than the widths of spiral arms in our Galaxy \citep[$\sim$ 200
  pc $-$ 400 pc,][]{rmb+14}. The Centaurus tangency region
($l\sim302^\circ-313^\circ$) seems complex, which is known for several
anomalies \citep[][]{ben08}. It has two distinct gas components (see
Fig.~\ref{all_plot} and Fig.~\ref{scutum}), one is near $l \sim
311^\circ$ and another is near $l \sim 306^\circ$. The longitude
difference for each component near the Centaurus tangency shown by
RRLs, CO, and HI is small, $\leq 1.4^\circ$ ($\sim$ 120
pc). Therefore, there is almost no obvious and ordered shift for
spiral arms traced by different gas tracers, i.e., RRLs, HII regions,
methanol masers, CO, dense gas regions, and HI.

However, remarkable difference is found between the arm tangencies
traced by old stars and gas (see Fig.~\ref{scutum}), which is about
$2.3^\circ-3.9^\circ$ ($\sim$ 310 pc $-$ 520 pc) for the northern part
of the Near 3 kpc Arm, $1.7^\circ-2.3^\circ$ in the Galactic longitude
($\sim$ 210 pc $-$ 290 pc) for the Scutum Arm, and about
$1.3^\circ-5.1^\circ$ ($\sim$ 120 pc $-$ 480 pc) for the Centaurus
Arm, which are comparable to the widths of spiral arms. In the
tangency regions of the northern part of the Near 3 kpc Arm and the
Scutum Arm, the stellar arms are exterior to the gas arms. As for the
Centaurus tangency, the offset direction of stellar arm relative to
gas arm is complex as it has two distinct gaseous components. The
relative strength of these two components shown by different gas
tracers (RRLs, CO, and HI) are not consistent (Fig.~\ref{scutum}). The
CO/HI integrated intensity for the component near $l\sim311^\circ$ is
stronger than that for the component near $l\sim305^\circ$
(Fig.~\ref{scutum}). By comparing the bump profiles derived from the
2MASS $K_s$ band data and the GLIMPSE 4.5$\mu$m data (see the right
panel in Fig.~\ref{scutum}), we found that the number increase of
infrared stars detected by GLIMPSE in the longitude range of $l\sim
309^\circ-314^\circ$ is much more prominent than that in the longitude
range of $l\sim 302^\circ-307^\circ$. Hence, the extinction caused by
interstellar dust is much more serious in the $l\sim
309^\circ-314^\circ$ directions, indicating a larger dust and gas
content. In addition, the literature mean values of the Centaurus
tangency \citep[see the summary given by][]{eg99,val14} are commonly
adopted as $\sim309^\circ$, closer to the gas component near $l \sim
311^\circ$. These features indicate that the bump component near
$l\sim311^\circ$ have a stronger gas concentration and larger gas
content, and the stellar Centaurus Arm is exterior to this
component. The shift of stellar arms exterior to CO gas arms increases
with the Galactocentric radius. Such offsets has also been observed in
some nearby galaxies \citep[e.g., M81, ][]{kkct08}.

\begin{figure}
\centering\includegraphics[width=0.48\textwidth]{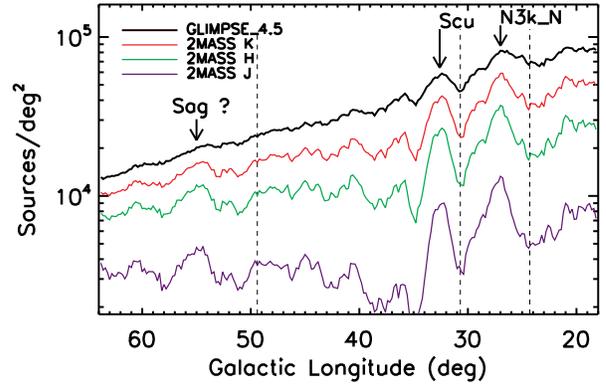}
\caption{Surface density of infrared stars as a function of the
  Galactic longitude, deduced from the survey data of GLIMPS 4.5
  $\mu$m, 2MASS $K_s$, $H$, and $J$ bands. The arm tangencies in
  longitudes for the northern part of the Near 3 kpc Arm (N3k$_{-}$N),
  the Scutum Arm (Scu), and the possible Sagittarius Arm (Sag) are
  indicated by arrows for the stellar components, and by dashed lines
  for the gaseous components.}
\label{sagittarius}
\end{figure}

Therefore, to identify the arm tangency features from the star count
plots, the possible offset between the stellar arms and gas arms
should be considered. We re-check the star count plots and found that
the stellar data from 2MASS {\it J}, {\it H}, and {\it K$_s$} bands
show a broad but small amplitude bump in the longitude plot near
Galactic longitude $\sim$ 55$^\circ$ (see
Fig.~\ref{sagittarius}). This bump, which has not been identified
\citep[e.g., see][]{dri00,cbm+09}, may be related to the tangency of
the Sagittarius Arm but shifted about 4.2$^\circ$ $-$ 5.8$^\circ$
($\sim$ 380 pc $-$ 530 pc) exterior to the gas arm. This direction
offset is consistent with the offsets for the Scutum Arm, the northern
part of the Near 3 kpc Arm, and the Centaurus Arm, and the offset
value ($\sim$ 4.2$^\circ$ $-$ 5.8$^\circ$) is larger but comparable to
that for the Centaurus Arm ($\sim$ $1.3^\circ-5.1^\circ$).

For the southern tangency of the Near 3 kpc Arm, the difference
between the arm tangency traced by stars and gas is not significant,
and around the tangency directions of the Carina Arm and the Norma
Arm, no obvious bumps can be found in the stellar data, so that for
these arms the possible displacements can not be properly assessed.

\begin{figure}
\centering\includegraphics[width=0.45\textwidth]{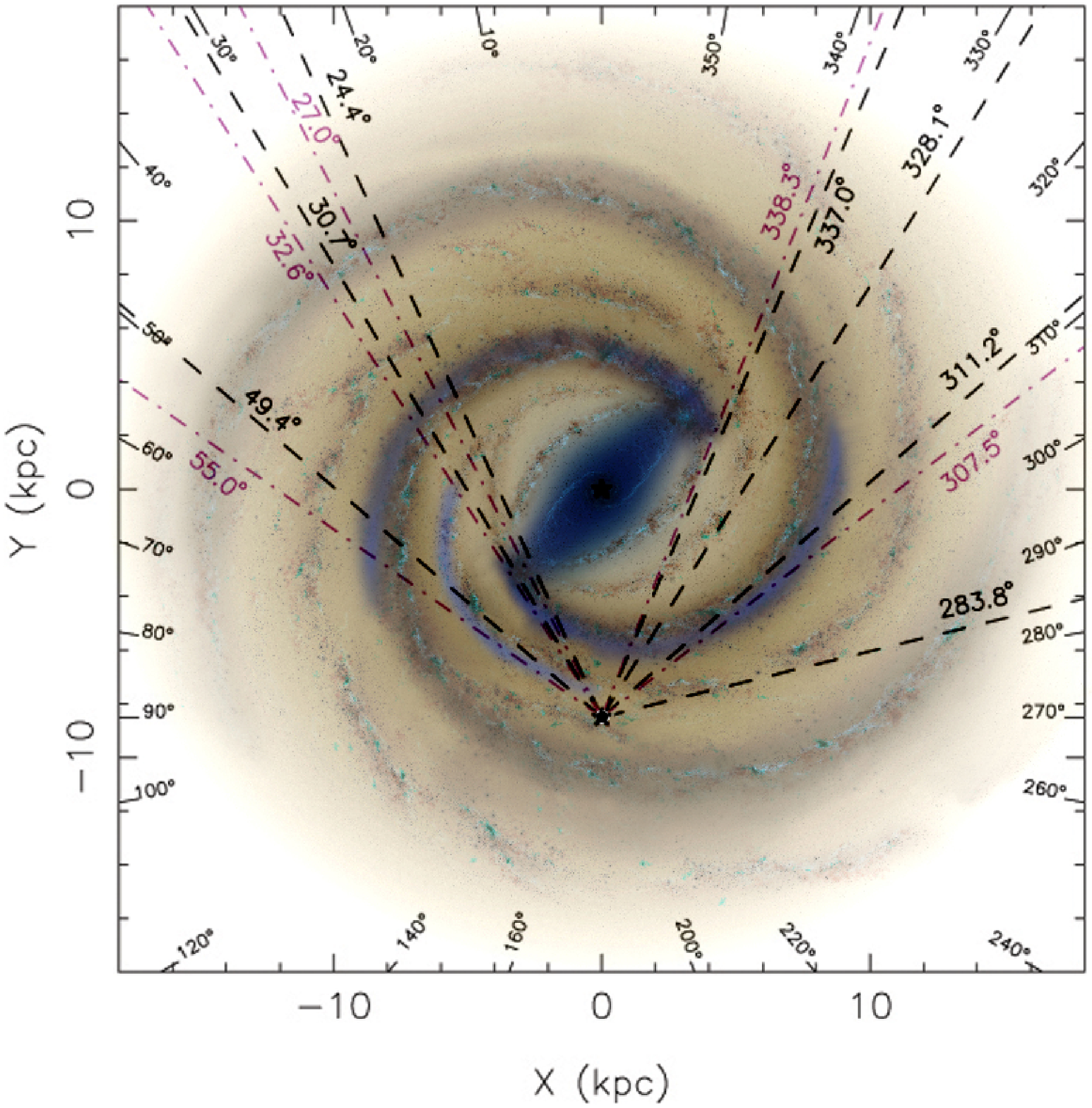}\\
\caption{A face-on image of the Milky Way modified from the artist's
  map (NASA/JPL-Caltech/R. Hurt) to match the observational properties
  summarized in this paper. The gaseous components in the Galactic
  plane are dominated by four arm segments \citep[see][]{hh14}. The
  stellar component is dominated by a two-arm spiral pattern but with
  a third weaker stellar arm of the Sagittarius. The newly derived arm
  tangencies in this paper are indicated by dashed lines for the gas
  arms, and dot-dashed lines for the stellar arms.}
\label{mw}
\end{figure}

\section{Discussions and Conclusions}

We reassess the Galactic longitude directions of spiral arm tangencies
in the Milky Way by using survey data for infrared stars and RRLs, HII
regions, 6.7 GHz methanol masers, CO, high density regions in gas, and
HI with a consistent definition for the tangencies. As verified by a
simple model, the directions of arm tangencies are very close to the
real density peaks of spiral arms near tangencies, but have a shift of
about 1$^\circ$ to the interior to the arm and the shift happens to
all kinds of spiral tracers.

Spiral arm tangency directions derived by different gas tracers, i.e.,
RRLs, HII regions, methanol masers, CO, dense gas, and HI, are almost
at the same longitudes for the Scutum-Centaurus Arm, the
Sagittarius-Carina Arm, the Norma Arm, and the Near 3 kpc
Arm. Therefore, we conclude that there is no obvious shift between the
gas arms, in spite of the gas phases, ionized or neutral or molecular.

We derived the tangency directions for the stellar arms, the
Scutum-Centaurus Arm, the Near 3 kpc Arm, and the Sagittarius Arm, and
obtain consistent values of arm tangencies by using stellar data of
GLIMPSE and 2MASS. However, we find remarkable difference between the
tangency directions of the stellar arms and the gas arms for the
Scutum-Centaurus Arm, the northern part of the Near 3 kpc Arm and
maybe the Sagittarius Arm, with a shift about 1.3$^\circ-$5.8$^\circ$
($\sim$ 120 pc $-$ 530 pc) in the Galactic longitude for different
arms, nearly about the widths of spiral arms in our Galaxy ($\sim$ 200
pc $-$ 400 pc).
The obvious offset between the stellar component near $l\sim 27^\circ$
and gas components near $l\sim 24^\circ$ for the Near 3 kpc Arm
deserves further studies since it could be related to a barred
potential \citep[e.g., $l\sim27^\circ$,][]{lhg01,bcb05}. The double
peaks for gas components near the Centaurus tangency are intriguing
and deserve for further investigations. The stellar overdensity near
$l\sim 55^\circ$ shown by using the 2MASS $J$, $H$ and $K_s$ band data
maybe related to the tangency of the stellar Sagittarius Arm, though
the feature is not obvious in the GLIMPSE data. The stellar
Sagittarius Arm may be much weaker than the stellar Scutum-Centaurus
Arm, and was smeared out in the GLIMPSE data. It is also possible that
the stellar overdensity shown in 2MASS data is caused by some nearby
stellar structures, and not related to the Sagittarius tangency. This
intriguing feature deserves further attention as it is important to
better understand the stellar density distribution in the Galactic
disk.

Based on observational properties of spiral structure we construct a
face-on image for the Milky Way as shown in Fig.~\ref{mw}, which shows
that the gaseous components are dominated by four {\it arm segments}
in the inner Galaxy regions \citep[e.g.,][]{gg76,rus03,hhs09,hh14},
and may extend to the far outer Galaxy
\citep[e.g.,][]{hh14,sxy+14}. The Local Arm is probably an branch of
the Perseus Arm \citep[e.g.,][]{xlr+13,hh14,bnh+14}. The distribution
of old stars in the Galactic plane is dominated by a two-arm spiral
pattern \citep[two major spiral arms: the Scutum-Centaurus Arm and the
  Perseus Arm, e.g.,][]{dri00,ds01,bcb05,cbm+09,fa12}. The Sagittarius
Arm may be the third but a weaker stellar arm. The tangency directions
for the density maximums of gas arm and stellar arm for the
Scutum-Centaurus Arm has a longitude difference as large as
1.6$^\circ-$ 4.7$^\circ$ ($\sim$ 150 pc $-$ 420 pc). In the inner
Galaxy, the stellar arms are shifted outwards with respect to the
gaseous arms traced by RRLs, CO and HI, not only for the
Scutum-Centaurus Arm, the northern part of the Near 3 kpc Arm, but
maybe also for the Sagittarius Arm.

Among three proposed mechanisms to produce spiral arms in galaxies,
only the quasi-stationary density wave theory predicts a spatial
displacement between the density peaks traced by stars and gas in
spiral arms, by assuming a constant pattern speed \citep[e.g.,][and
  see Fig.~\ref{show}]{rob69,db14}. Our data analyses show the shift
between the directions for the density maximums of stars and the
intensity peaks for gas emission in the Scutum-Centaurus Arm near the
tangencies, which is evidence for the quasi-stationary density wave in
the Milky Way. Considering the symmetry of the Galaxy spiral structure
proposed in literature \citep[e.g.,][]{dam13}, such an offset between
gas and old stars is also expected in the Perseus Arm (see
Fig.~\ref{mw}), and recently discussed by \citet[][]{mgf15} using the
stellar overdensity and the interstellar visual absorption in the
anticenter direction. The accurate measurements of the parallax/proper
motion for stars by Gaia \citep[e.g.,][]{rlr+12,khg+15}, and for
high-mass star forming regions by the Bar and Spiral Structure Legacy
(BeSSel) Survey\footnote{http://bessel.vlbi-astrometry.org} and the
Japanese VLBI Exploration of Radio Astrometry
(VERA)\footnote{http://veraserver.mtk.nao.ac.jp} will provide critical
tests in the near future. In addition, no {\it general} spatial
ordering can be reliably deduced for the multiphase gas
arms. Considering the multiphase properties of interstellar medium and
the stellar feedback \citep[e.g.,][]{wad08}, the response of gas to
the Galaxy potential
probably highly complex, making it difficult to verify the possible
spatial ordering for multiphase gas arms in observations at present.

\section*{Acknowledgements}
We appreciate the anonymous referee for the instructive comments which
help us to improve the paper. The authors are supported by the
Strategic Priority Research Program ``The Emergence of Cosmological
Structures" of the Chinese Academy of Sciences, Grant No. XDB09010200,
and the National Natural Science Foundation (NNSF) of China
No. 11473034. L.G.H. is also supported by the Young Researcher Grant
of National Astronomical Observatories, Chinese Academy of Sciences.

\footnotesize

\bibliographystyle{aa} 
\bibliography{nh3}
\normalsize

\appendix

\end{document}